\documentclass[letterpaper]{natureprintstyle}
\usepackage{amsfonts}
\usepackage{eurosym}
\usepackage{graphicx,epsfig}
\usepackage{amsmath,amssymb,bbm}
\usepackage{color}
\usepackage[usenames,dvipsnames]{xcolor}
\usepackage[colorlinks=true,linkcolor=Red,citecolor=Green,linktoc=page]{hyperref}
\usepackage{multirow}
\usepackage[varg]{txfonts}
\usepackage{ulem}
\usepackage{fancyhdr}
\usepackage{mathtools}
\usepackage[caption=false]{subfig}
\usepackage{balance}
\usepackage{float}
\usepackage{flushend}

\def\subinrm#1{\sb{\rm#1}}
{\catcode`\_=13 \global\let_=\subinrm}
\mathcode`_="8000
\def\upsubscripts{\catcode`\_=12 }

\upsubscripts

\newcommand{\sclprod}[2]{(#1 \cdot #2)}
\newcommand{\vecprod}[2]{[#1 \times #2]}

\graphicspath{ {Figures/} }

\begin{document}

\title{Time-dependent exchange  creates the time-frustrated state of matter}
\author{V.\,E.\,Valiulin $^{1,2}$,
	N.\,M.\,Chtchelkatchev\,$^{1}$,
	A.\,V.\,Mikheyenkov\,$^{1,2}$
	\& 	V.\,M.\,Vinokur\,$^{3,4,\ast}$.
}

\date{Magnetic systems governed by exchange interactions between magnetic moments harbor frustration that leads to ground state degeneracy and results in the new topological state often referred to as a frustrated state of matter (FSM). The frustration in the commonly discussed magnetic systems has a spatial origin. Here we demonstrate that an array of nanomagnets coupled by the real retarded exchange interactions develops a new state of matter, time frustrated matter (TFM). In a spin system with the time-dependent retarded exchange interaction, a single spin-flip influences other spins not instantly but after some delay. This implies that the sign of the exchange interaction changes, leading to either ferro- or antiferromagnetic interaction, depends on time. As a result, the system's temporal evolution is essentially non-Markovian. The emerging competition between different magnetic orders leads to a new kind of time-core frustration. To establish this paradigmatic shift, we focus on the exemplary system, a granular multiferroic, where the exchange transferring medium has a pronounced frequency dispersion and hence develops the TFM.}
\maketitle
\upsubscripts
\thispagestyle{fancy}
\lfoot{\parbox{\textwidth}{ \vspace{0.3cm}
 \rule{\textwidth}{0.2pt}
\hspace{-0.2cm} \textsf{\scalefont{0.80}
    $^1$Vereshchagin Institute of High Pressure Physics, Russian Academy of Sciences, 108840 Troitsk, Moscow, Russia;
    $^2$Moscow Institute of Physics and Technology, 141701, Dolgoprudny, Russia;
    $^3$Terra Quantum AG, St. Gallerstrasse 16A, CH-9400 Rorschach, Switzerland;
    $^4$Physics Department,
    City College of the City University of New York,
    160 Convent Ave, New York, NY 10031, USA;
    $^*$correspondence to be sent to vmvinokour@gmail.com
}
\vspace{-0.2cm}
\begin{center}{\scalefont{0.87} \thepage}\end{center}}} \cfoot{}


Macroscopic magnetism forms due to microscopic exchange interactions\,\cite{Magnetism2006,Magnetism2007,Magnetism2021}.
The exchange interaction is of the quantum mechanical origin and stems from the intertwined effect of the Coulomb interaction and Pauli exclusion principle governing the behavior of indistinguishable fermions with overlapping wave functions. Remarkably, a rich lore narrating how the geometric frustration developing from the exchange effects leads to degeneracies and the emergent FSM, neglects the temporal component: local magnetic moments are supposed to instantly interact with each other without delays\,\cite{Manipa19_N}. However, the flip of electron spin in an atom in a crystal implies a rearrangement of electron density distribution in space, which, in turn, affects the strength of interactions between the atom with its neighbors. The rearrangement of electron clouds occurs at optical frequencies ($ \sim500~\mathrm{THz}$) and so  with characteristic time scales $\sim0.01~\mathrm{ps} - 0.1~\mathrm{ps}$, while the relaxation of the atomic position caused by the spin flip occurs at phonon frequencies ($\sim1~\mathrm{THz}$) and so picosecond time scales\,\cite{Kirilyuk10_RMP}. As a result, the spin exchange interaction in solids should be in general nonlocal in time and has the time delayed (retarded) nature.

There is an abundance of new functional materials, like granular multiferroics \cite{Spaldi19_NM,Spaldi17_NRM}, where interactions occur not directly but through the mediating active dielectric or ferroelectric environment. In such materials\,\cite{Dong15_AP,Fiebig16_NRM,Udalov14_PRB,Udalov14_PRBb} the retardation effects are relatively large and cannot be ignored. There, the polarization, $\mathbf P$, of a ferroelectric manifests retarded response to the electric filed $\mathbf E$, hence $\mathbf P(t)=\int \hat\alpha(t-t')\mathbf E(t')dt'$, where $\hat\alpha$ is the polarizability tensor~\cite{Maity20_PSSB,Ma20_PSSR} in a linear response approximation. The superexchange interaction of magnetic moments in granular multiferroics\,\cite{Fedoro14_PRBa}, where electric and magnetic degrees of freedom mutually influence each other, acquires retardation as well. In our work we reveal the time retardation of the exchange interaction and investigate the time-frustrated state of matter emerging due to this retardation in an array of magnetic moments immersed into the ferrolectric environment.

Relaxation of the exchange has recently been intensely studied, both experimentally and theoretically \cite{Secchi13_APN,Claass17_NC,Mentin17_JPCM,Liu18_PRL,Chaudh19_PRB,Ke20_PRR,Mikhay20_PRL,Losada19_PRB,Ron20_PRL}. One of the most discussed examples has been relaxation of the system after an instantaneous external action of, for example, laser irradiation finite-duration pulse. Here, we address a different situation where the retardation emerges due to internal properties of the system. The resulting relaxation manifests a variety of the non-trivial effects appearing even without the external finite-time impacts. We are confident that our finding on the retarded nature of exchange would contribute to investigations of relaxation processes in the time-dependent exchange.

\section*{The model}
\begin{figure}[t!]
\centering
	\includegraphics[width=0.48\columnwidth]{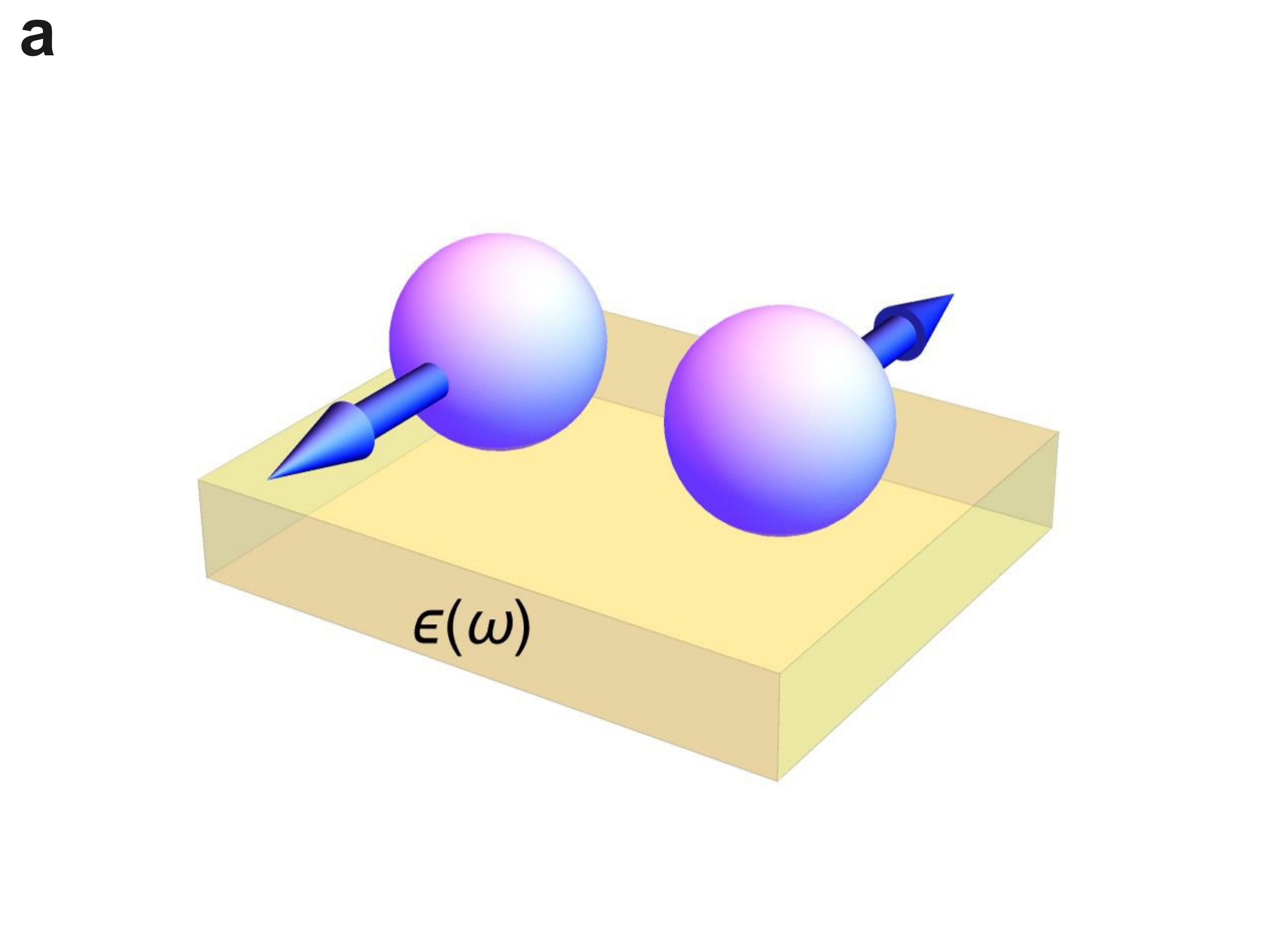}
	\includegraphics[width=0.48\columnwidth]{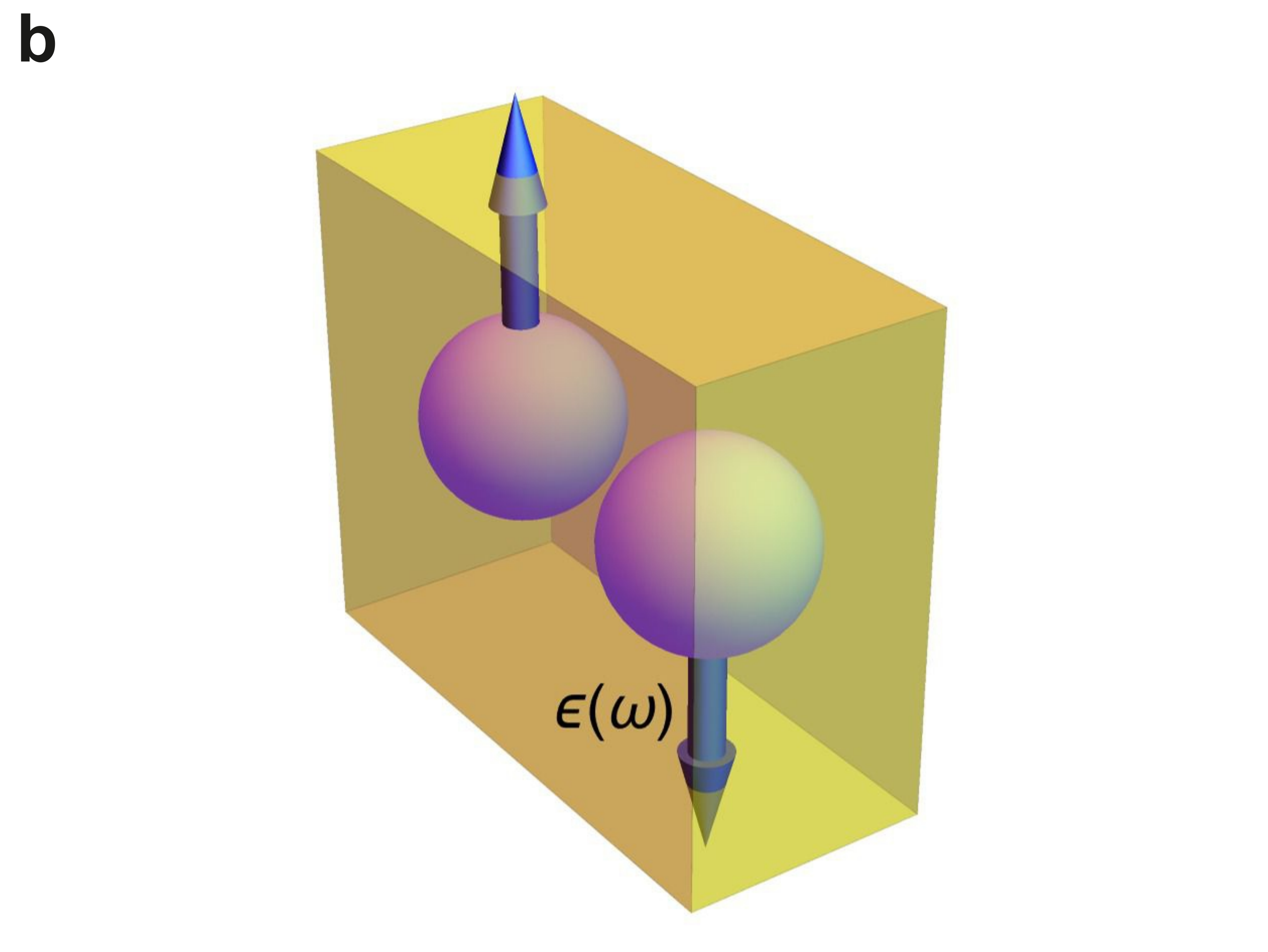}
	\caption{\textbf{The origin of the retarded superexchange spin-spin interaction in a granular multiferroic.} The  wave functions of electrons located at adjacent magnetic metallic granules (spheres) overlap in the ferroelectric medium (yellow semitransparent environment) to form the exchange integral $J$  depending on the frequency through the ferroelectric permittivity $\varepsilon(\omega)$. In time representation, $J(t)$ is the time-retarded quantity. {\bf a}, Granules are superimposed over the dielectric substrate. {\bf b}, Granules are immersed into the ferroelectric environment.
	}
	\label{Fig1}
\end{figure}

~~\\
To reveal how the exchange retardation results in the TFM we focus first on an elemental building block of a granular-multiferroic, two adjacent magnetic granules interacting via a ferroelectric medium as schematically shown in Fig.\,\ref{Fig1}. Figure\,\ref{Fig1}a displaces two metallic granules carrying the opposite magnetic moments disposed over the ferroelectric substrate and Fig.\,\ref{Fig1}b presents the same magnetic moments immersed into a ferroelectric medium. Since the dielectric constant of a ferroelectric environment typically has significant frequency dispersion, the retardation effects are inevitable. Indeed, in a simplest approximation taking into account the dielectric screening of the Coulomb interaction, one finds, following\,\cite{Udalov14_PRB,Udalov14_PRBa,Udalov14_PRBb}, that the dielectric constant appears in the effective exchange between two magnetic moments as
\begin{equation}
	\label{eq_1}
	J \! \sim \!  \sum_{a,b} \! \int d\mathbf{r_{1}} d\mathbf{r_{2}} \Psi _{a}^{*} (\mathbf{r}_{2} )\Psi _{b}^{*} (\mathbf{r}_{1} )\frac{e^{2} }{\varepsilon \, |\mathbf{r}_{1} -\mathbf{r}_{2} |} \Psi _{a} (\mathbf{r}_{1} )\Psi _{b} (\mathbf{r}_{2} )\,,
\end{equation}
where the sum is taken over the electron wave functions of each granule, and $\Psi _{a,b} (\mathbf{r}_{1,2} )$ stand for the undisturbed electron wave functions.
More precise calculation requires including the effect of the environment on the wave functions\,\cite{Udalov14_PRBa} and also accounting for the spacial dispersion effects.
Yet, even in this first approximation, the frequency dispersion of the exchange integral $J(\omega)$ arises due to dispersion of $\varepsilon (\omega)$ the behavior of which is straightforwardly  related to the dielectric permittivity tensor of the ferroelectric environment\,\cite{Udalov14_PRB,Udalov14_PRBb,Fedoro14_PRBa,Udalov14_PRBa,Fedoro15_PRB}.
Accordingly, we arrive at the model of a magnetic system with the effectively delayed exchange. In a temporal representation, this implies the delay in the interaction of magnetic moments:
$\int J_{12}(t-t')\,{\mathbf{m}}_{1}(t)\,{\mathbf{m}}_{2}(t')dt'$, where $J_{12}(t-t')$ is the Fourier transform of $J(\omega)$. It is essential that $J_{12}(t-t')$ is purely retarded, that is $J_{12}(t-t') =0$, if $t<t'$, so the causality is fulfilled.

In typical ferroelectrics (e.g., such as barium titanate (BTO) and lead zirconate titanate (PZT)), $\varepsilon(\omega)$ is large at low frequencies, $\varepsilon\,(\omega$$=$$0)= \varepsilon_0$$\gtrsim$$1000$, and
is of the order of unity for large frequencies, $\varepsilon\,(\omega$$=$$\infty)= \varepsilon_{\infty}$$\simeq$$1$.
The frequency threshold is set by the phonon frequency which usually does not exceed\,1\,THz.
Consequently, for small frequencies, we may treat $J(\omega)$ as vanishing, while at large frequencies, $J(\omega)$ tends to finite values.

Accordingly, we put
\begin{equation} \label{J_w0}
	J_{12}\,(\omega=0) = \int_0^\infty J_{12}(t)dt = 0,
\end{equation}
implying that the function $J(t)$ is alternating in sign with time. Hence we arrive at the ``time frustration'' of the exchange interaction.

Let us consider now two adjacent magnetic moments $\mathbf{m}_{1}$, $\mathbf{m}_{2}$. The delay in the interaction implies a nonequilibrium regime at finite times, while due to the nonzero damping the magnetic moments assume stationary values, $\mathbf{m}_{1,2}^{(\infty)}$, at $t$$\to$$\infty$. The final magnetic state is to be derived from the energy considerations using the effective exchange Hamiltonian
$H=J_{12}(\omega$$=$$0)\, \mathbf{m}_{1}^{(\infty)} \mathbf{m}_{2}^{(\infty)}$.
The mutual orientation of magnetic moments is to be found by investigating the magnetization time evolution from the starting point to $t\to\infty$.

The magnetic granules are supposed to be semiclassical, hence granule's magnetization should obey the non-local in time Landau-Lifshitz-Gilbert (LLG) equation. We consider an array of localized magnetic moments\,satisfying $|\mathbf{m}_i(t)|$$=$$1$ condition. The\,equation\,of\,motion\,for\,$i$-th moment\,is
\begin{equation} \label{M_i}
	\mathbf{\dot{m}}_{i}(t) = -{\gamma}[\mathbf{m}_i(t) \times \mathbf{h}_i^{\rm{eff}}(t)]
	- \lambda\gamma[\mathbf{m}_i(t) \times [\mathbf{m}_i(t) \times \mathbf{h}_i^{\rm{eff}}(t)]],
\end{equation}
were, as usual, $\gamma$ is the gyromagnetic ratio and $\lambda$ is the damping parameter. Here $\mathbf{h}_i^{\rm{eff}}$ is an effective Weiss field
at the i-th cite defined as
\begin{equation} \label{H_eff}
	\mathbf{h}^{\rm{eff}}_{\mathbf{i}} (t) = \sum_{\mathbf{j}} \int_{-\infty}^{t} J_{\mathbf{ij}}(t-\tau) \mathbf{m}_{\mathbf{j}}(\tau) d\tau + \mathbf{h}^{\rm{ext}}(t),
\end{equation}
where $\mathbf{h}^{\rm{ext}}(t)$ is the weak external magnetic field,
and the sum runs over the nearest neighbors to the site $\mathbf{i}$ (we account for only the nearest-neighbor exchange interactions). Exchange integrals $J_{\mathbf{ij}}(t)$ are time-dependent and preserve the causality.
An instant interaction
$J_{\mathbf{ij}}(t$$-$$t')$$=$$\delta(t-t'-0)J_{\mathbf{ij}}$
corresponds to the standard LLG equation~\cite{landau2013electrodynamics,Nikolic2019PRB} implying that
the exchange energy assumes the usual form,
$-1/2 \sum_{\mathbf{i},\mathbf{j}} J_{\mathbf{ij}}\ \mathbf{m_i}\, \mathbf{m_j}$.

As usual, in the dimensionless equation (\ref{M_i}) and (\ref{H_eff}) the magnetic moment and all the fields are normalized by the magnetic moment's length (saturation magnetization per site) $M_s$, the time is measured in units of $(\gamma_0 M_s)^{-1}$,
$\gamma_0$ being the gyromagnetic ratio. This time scale has the order of $10^{12} s^{-1}$ (that corresponds to $\mathrm{1~THz}$) with $\gamma_0 = 1.39 \times 10^6 \mathrm{rad/s(A/m)}$ and $M_s \sim 10^6 \mathrm{A/m}$ (for $\mathrm{FePt}$) \cite{Spaldi19_NM,John17_SR,Liu17_APL,Jin18_PRE}.

\section*{Stationary solutions to the LLG equation}~~\\
Let us consider a stationary solution to Eq.\,(\ref{M_i}) in which we hereafter set $\gamma = 1$ for simplicity.
We assume that $\mathbf{m}_{i}(t)$$=$$\mathbf{m}^{0}_{i}$ is a set of stationary solutions to Eq.\,(\ref{M_i}) in the absence of the external field, $\mathbf{h}^{\rm{ext}}(t) = 0$. For better visibility we further simplify the notations and write:
\begin{eqnarray}
	\label{simp}
	J_0 = \int_{-\infty}^{t} J(t-\tau) d\tau  \\
	\mathbf{h}_i^{0} = \sum_{NN} J_0 \mathbf{m}^{0}_{NN}\\
	\mathbf{b}_{i}^{0} = \vecprod{\mathbf{m}^{0}_i}{\mathbf{h}_i^{0}}\,,
\end{eqnarray}
where $\sum_{NN}$ stands for the sum over the nearest neighbors. In this transparent case, the LLG equation assumes the form
\begin{equation}
	\label{LL_s_0}
	\mathbf{\dot{m}}^{0}_{i}(t)=0=-\mathbf{b}_{i}^{0}
	-\lambda \vecprod{\mathbf{m}^{0}_{i}}{\mathbf{b}_{i}^{0}},
\end{equation}
which requires that $\mathbf{b}_{i}^{0} = 0$.
Stationary solutions would realize  for either
\begin{itemize}
\item Frustrated exchange with
$J_{0}$$=$$\int_{-\infty }^{t \to \infty}J(t-\tau)d\tau$$=$$0$
implying $\mathbf{h}_{i}^{0}\,\mathbf{\equiv 0}$,
hence $\mathbf{b}_{i}^{0}\,\mathbf{\equiv 0}$.
Therefore, any magnetic configuration formally assumes a stationary solution. We address further the important question whether these solutions are stable with respect to small perturbations like noise or an external field, and show that only particular stationary configurations are stable.
\end{itemize}
or for
\begin{itemize}
\item A non-frustrated exchange with
$J_{0}$$=$$\int_{-\infty}^{t \to \infty}J(t-\tau )\,d\tau \neq 0$, where
we have a condition $\mathbf{b}_{i}^{0}\,=
J_{0} \vecprod{\mathbf{m}_{i}}{\sum_{NN} \mathbf{m_{NN}}} = 0$\,,
which is satisfied for common FM, AFM structures and for collinear configurations for which
$\sum_{NN} \mathbf{m_{NN}}$$=$$0$, e.g., stripe structures in square lattice.
\end{itemize}
We see that for stationary solutions, frustrated and non-frustrated cases differ qualitatively. Hereafter we focus on a dynamically frustrated case.

\section*{Time-dependent magnetic moments}

\subsection{General equations}

Let us derive the adjustments to the stationary solution discussed above arising due to time dependence of the magnetic moments. We consider the time-dependent part of the $i$-th magnetic moment $\mathbf{m}^{\delta}_{i}(t)$ being small and write magnetization as
\begin{equation}
\label{adj_def}
  \mathbf{m}_{i}(t)=\mathbf{m}^{0}_{i} + \mathbf{m}^{\delta}_{i}(t)\,.
\end{equation}
In the corresponding linear approximation, the Landau-Lifshitz-Gilbert equation for $\mathbf{m}^{\delta}_{i}(t)$ in the frustrated case assumes the form, see Methods
\begin{equation} \label{LL_frustr}
\dot{\mathbf{m}}_{i}^{\delta }(t)=-[{\mathbf{m}_{i}^{0}\times } {\mathbf{h}_{i}^{\delta }(t)]}-\lambda \lbrack {\mathbf{m}_{i}^{0}\times \lbrack {\mathbf{m}_{i}^{0}\times }{\mathbf{h}_{i}^{\delta }(t)]}]}\,.
\end{equation}

To find the analytical solution to Eq.\,(\ref{LL_frustr}) we take its Fourier transform and obtain, see Methods,
\begin{equation} \label{LL_omega1}
i\omega \mathbf{m}_{i}^{\delta }(\omega )=-[{\mathbf{m}_{i}^{0}\times }{%
\mathbf{h}_{i}^{\delta }(\omega )]}-\lambda \mathbf{m}_{i}^{0}({\mathbf{m}%
_{i}^{0}}\mathbf{\cdot }{\mathbf{h}_{i}^{\delta }(\omega ))}+\lambda {%
\mathbf{h}_{i}^{\delta }}(\omega )\,.
\end{equation}

\subsection{Two-site cluster with the frustrated exchange}
The above general reasoning holds for any arbitrary regular magnetic structure and, in particular, is not restricted to systems subject to nearest neighbor interactions constraint.
To illustrate how the formation of the time-frustrated state occurs, we consider the simplest particular system,
two interacting magnetic moments $\mathbf{m}_{1}$ and $\mathbf{m}_{2}$.
To further simplify the problem, we analyze collinear stationary configurations,
FM with ${\mathbf{m}_{1}^{0} = \mathbf{m}_{2}^{0}\parallel z}$
and AFM with ${\mathbf{m}_{1}^{0} = - \mathbf{m}_{2}^{0}\parallel z}$
(hereafter we use
$\mathbf{m}_{i}^{0}\cdot \mathbf{m}_{i}^{0}=1,\ i=1,2$
and $\mathbf{m}_{1}^{0}\times \mathbf{m}_{2}^{0} = 0$).

\begin{figure}[h] \centering
  \includegraphics[width=0.75\columnwidth]{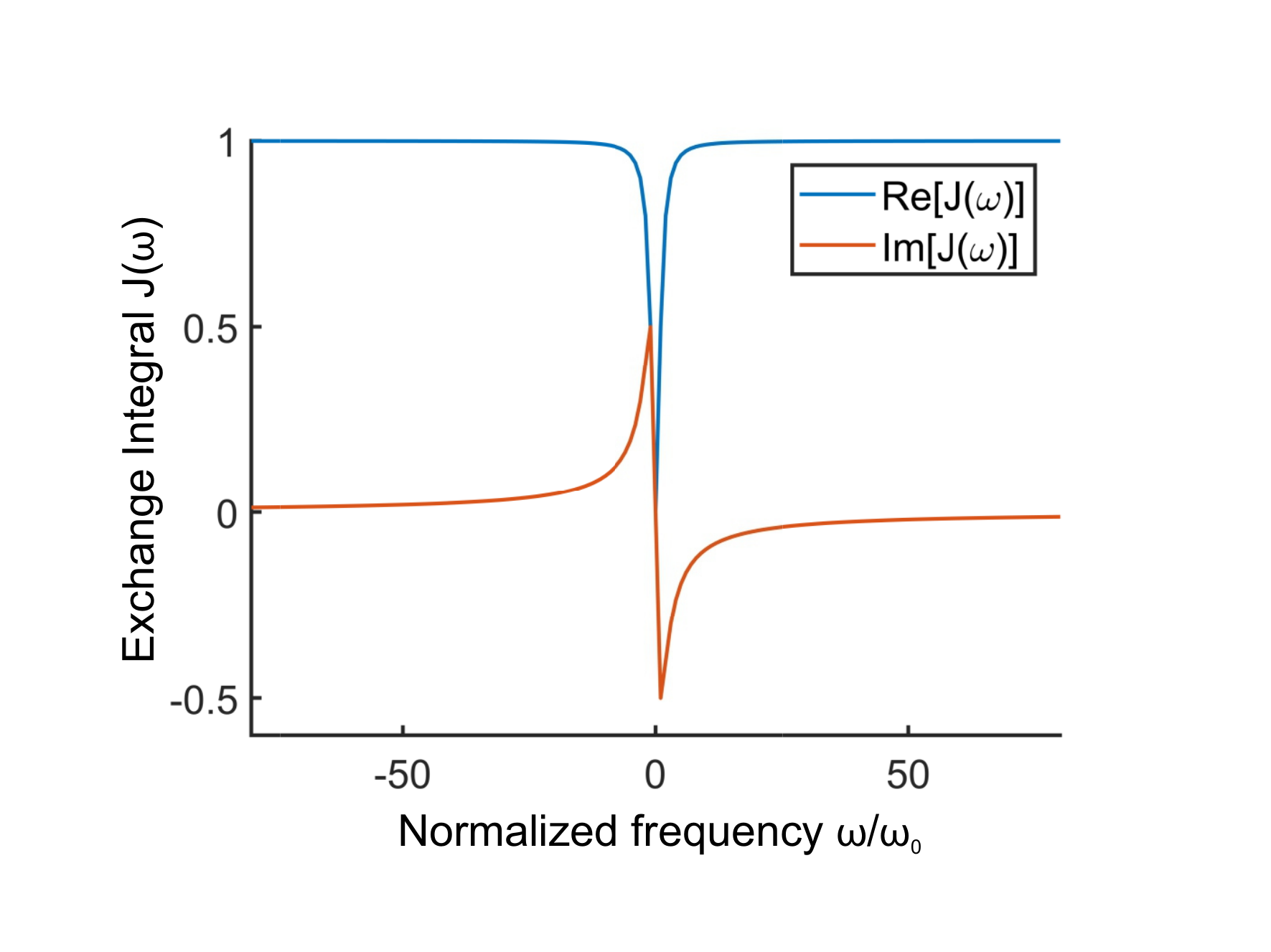}
	\caption{\textbf{Real and imaginary parts of the adopted time frustrated exchange potential} ${J}(\omega) = G\, {\omega}/({\omega + i\omega_0})$. Here the reference frame is $J(\omega \to \infty) = G =1$.}
	\label{Fig_Re_Im}
\end{figure}

Taking the simplest form of the exchange satisfying all the above-defined conditions
\begin{equation} \label{Jt_2}
J(t) = G (\delta(t) - \omega_0 e^{-\omega_0 |t|})\,
\end{equation}
where $\delta(t)$ is the Dirac delta function using its Fourier transform,
\begin{equation} \label{J_simp_2}
{J}(\omega) = G \frac{\omega}{\omega + i\omega_0}\,,
\end{equation}
see Fig.\,\ref{Fig_Re_Im}, which preserves the causality as the pole is in the lower half of the $\omega$-plane, and displays a reasonable asymptotic behavior:
$J(\omega$$=$$0)$$=$$0$,
$J(\omega$$\to$$\infty)$$\to \textrm{const}$.
The characteristic value $\omega_0 \sim 1$ with the accepted time scale corresponds to $THz$ phonon frequency region of the common granular multiferroics, e.g.
$\mathrm{Pr_{1-x}SrMnO_3/LuMnO_3}$, $\mathrm{NaNO_2/porous\ glass}$,
$\mathrm{SrTiO_3/rutile}$ $\mathrm{SrTiO_3/Teflon}$,
$\mathrm{La_{5/8}BaCa_{3/8}MnO_3/LuMnO_3}$
\cite{Spaldi19_NM,Akbash15_JMMM,Colla99_ASS,Park04_PRL,Mandal06_PRB}

To simplify further notations, we set it in that the magnetic moments and energy are properly normalized and are measured in dimensionless units. Thus, $G$, $\gamma$ and $\lambda$ also become dimensionless.
Now we find the stability conditions ensuring the stationary solutions. For the FM case, $m_{1}^{0}=m_{2}^{0}=+1$, we obtain
the stability condition as $G\lambda <\omega_0 $.
For the AFM configuration, $m_{1}^{0} = - m_{2}^{0} = +1$, and
the resulting stability condition is
${G}\sqrt{1+\lambda^{2}} < \omega_0 $.

Having established the ranges of stability within the linear approximation, let us turn to detailed investigating the time evolution of our system.
The zero-frequency limit was discussed above. At high frequencies the ferroelectric degrees of freedom are frozen and the exchange is provided by the conventional electron clouds overlapping.

\begin{figure}[t] \centering
	\includegraphics[width=0.75\columnwidth]{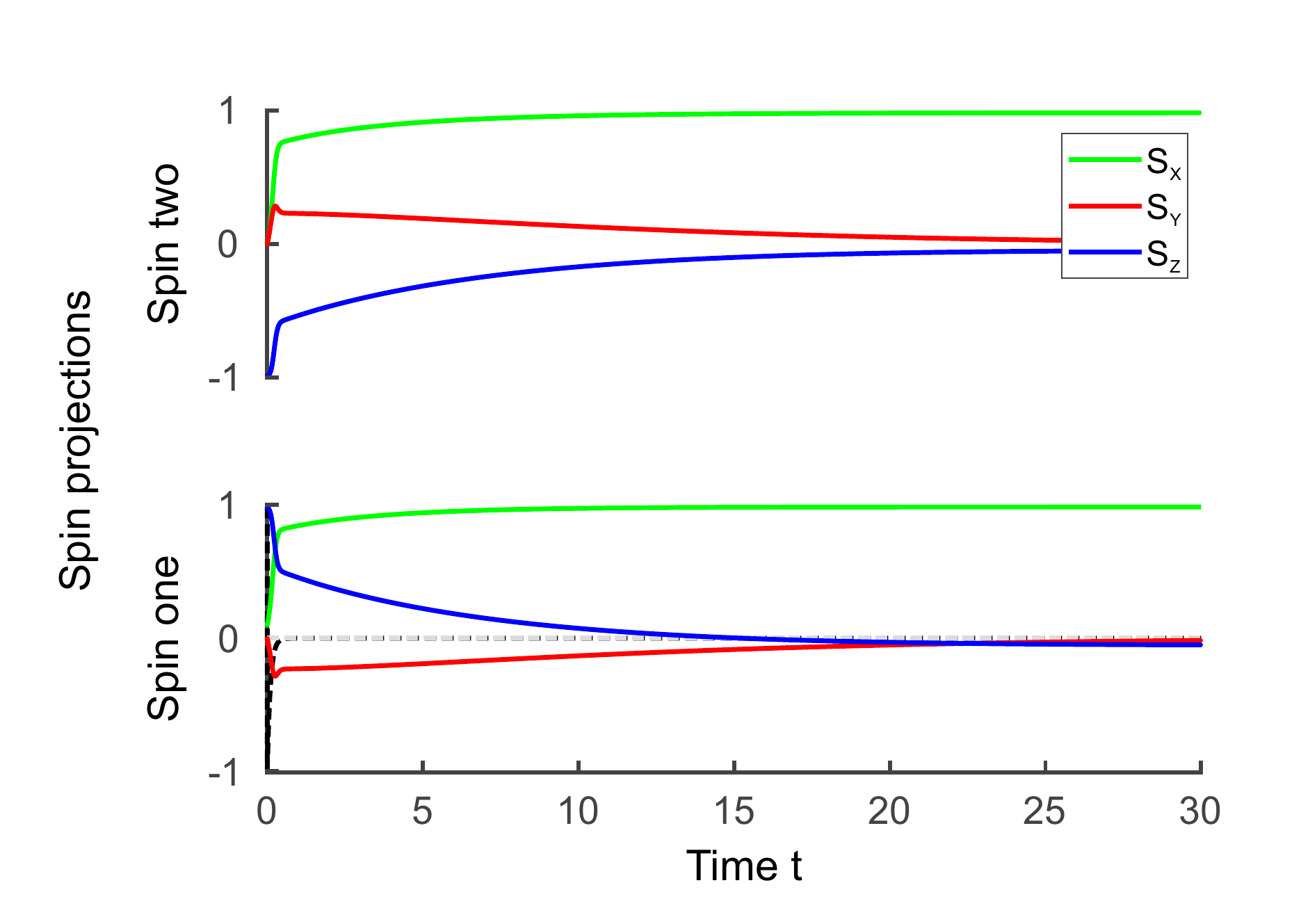}\\ 
	\includegraphics[width=0.75\columnwidth]{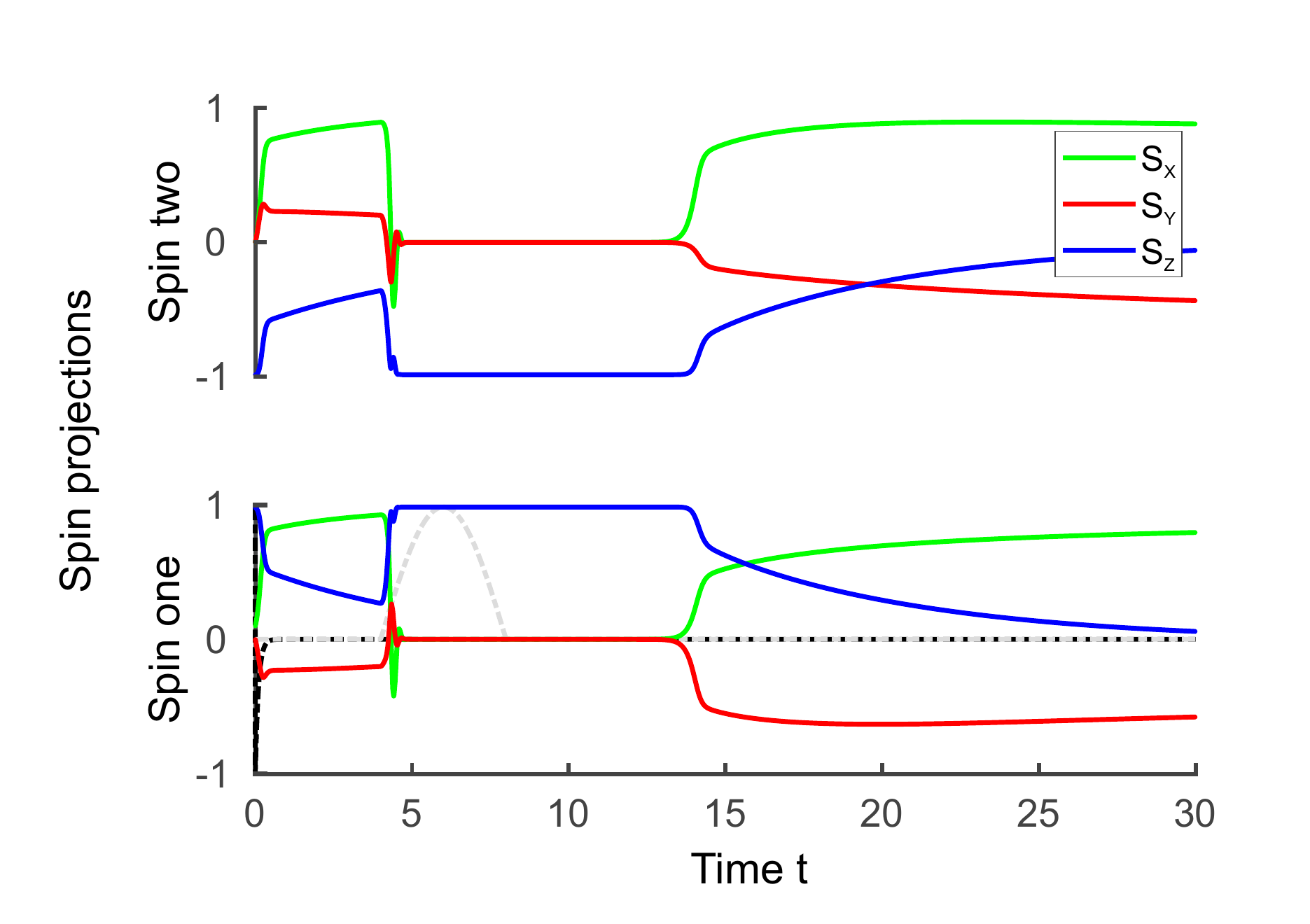}
	\caption{\textbf{Evolution of the two-site state with the time-retarded exchange interaction 
			${J}(\omega) = G\, {\omega}/({\omega + i\omega_0})$.} \textbf{a:} The evolution in the absence of anisotropy. Magnetic moment projections for both sites are depicted by different (blue, red and yellow) colors. It is seen that non-Markovian time-frustrated exchange transforms AFM initial state into stable FM state.
	\textbf{b:} The effect of perturbation.  The perturbation in the form of magnetic pulse (grey dash-dotted line) converts the stable FM state into long-living AFM one. Then the stable FM restores. The Landau-Lifshitz-Gilbert equation parameters for both panels are $\gamma = 1$, $\lambda = 1$. The retarded exchange parameters are $G = 10$, $\omega_0 = 10$. Black dash-dotted lines (lower half of each pair) depict the exchange potential, the retardation is becoming almost invisible in the adopted time scale. Spin projections and exchange scales differ.
		The initial state on both panels is the slightly disturbed AFM (one magnetic moment infinitesimally rotated from the pure AFM). Note that the moderate noise does not qualitatively affect the described process.
	}
	\label{Fig3}
\end{figure}

The time evolution appears radically different in the isotropic case and in the presence of the even weak uniaxial anisotropy.
In the isotropic case, the asymptotic, $t \to \infty$ state of magnetic moments, either in the FM or AFM case, is defined by the exchange potential parameters, mostly by its $\delta$-part. In the anisotropic case either the $\delta$-part or the exponential part of the time-depending exchange (\ref{Jt_2}) dominates the system's behavior.
The results of the numerical calculations of the time evolution are displayed in Fig.\,\ref{Fig3}.
The panel Fig.\,\ref{Fig3}a shows that the evolution of the isotropic system with the time-dependent exchange results in the FM state at $t\to\infty$, as it is clearly seen from the evolution of magnetic moments projections. Remarkably, although perturbing the system by
the half-sinusoidal pulse of the external magnetic field switches the FM state into the AFM one, see the panel Fig.\,\ref{Fig3}b, this AFM state lives only for some finite time, and then the system returns back to the FM state.

This whole evolution picture is noise-resistant, it does not transform under the delta-correlated noise with the amplitude $|\mathbf{h_{noise}}|$ small relative to effective field $|\mathbf{h}_{noise}| \ll |\mathbf{h}^{\rm{ext}}(t)|$.
The external perturbation in the form of the sequence of the alternating pulses successively converts FM to AFM and vice versa, see Supplementary Information (SI).

\begin{figure*}[tbp]
\centering
	\includegraphics[width=0.9\columnwidth]{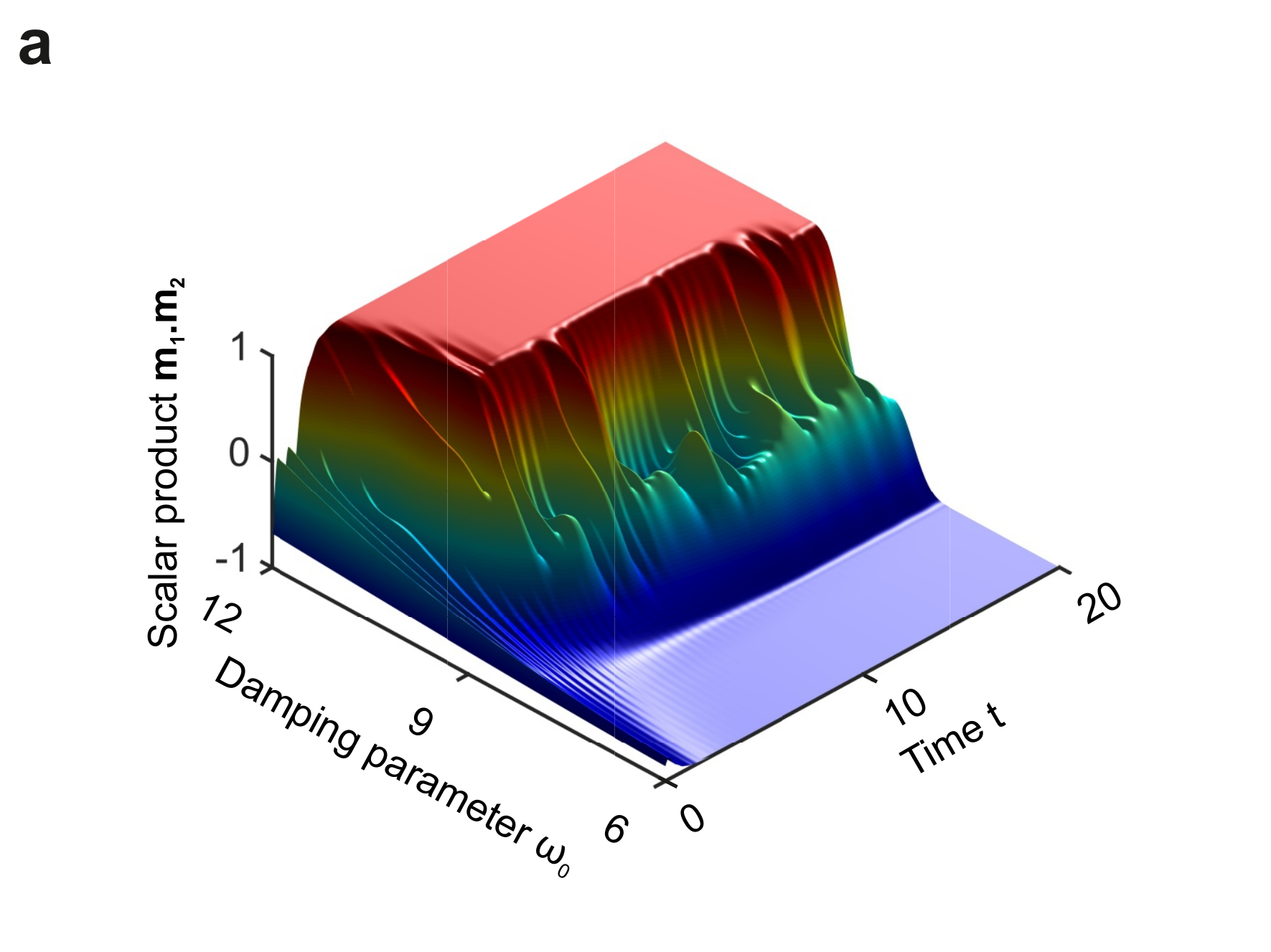} \hspace*{4pt}
    \includegraphics[width=0.9\columnwidth]{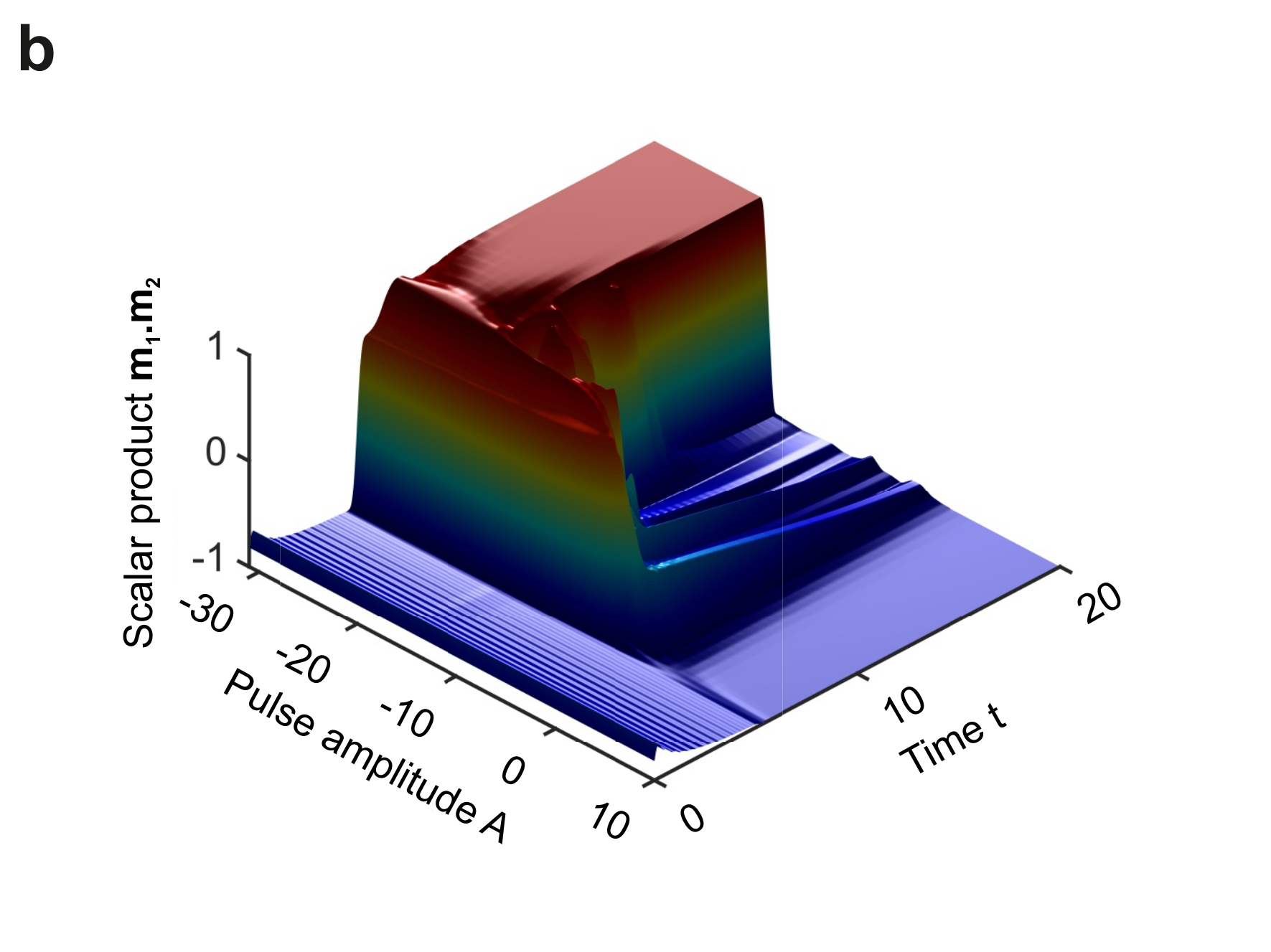}
	\caption{\textbf{Control of the final stable state by the damping parameter or by the external pulse: the magnetic moments scalar product $\mathbf{{m_1}\cdot {m_1}}$ is depicted.}
	 \textbf{a,} In the presence of weak anisotropy, the initial state transforms into different final stable states depending on damping parameter $\omega_0$.
If $\omega_0 \lesssim\omega^{\ast}\simeq 7.5$ (relatively slow retardation), the final stable state is the FM, for $\omega_0 \gtrsim\omega^{\ast\ast}\simeq 10$ (relatively fast retardation), the final stable state is the AFM. The initial state is the slightly disturbed AFM (one magnetic moment infinitesimaly tilted away from purely AFM arrangement). The asymptotic, $t \gg 1$, picture does not depend upon the initial state for $\omega_0\not\in(\omega^{\ast},\omega^{\ast\ast})$. The Landau-Lifshitz-Gilbert equation parameters are $\gamma = 1$, $\lambda = 1$, anisotropy parameter $\rho = 5$, the retarded exchange amplitude is $G = 10$.
\textbf{b,}	In the presence of weak anisotropy, the perturbation in the form of the magnetic pulse (pulse being the half sinusoidal, $A\sin(\alpha t)$) allows to control the final stable sate.
For $A \lesssim A^{\ast} -15$ (negative pulse) the final stable state is the FM, for $A \gtrsim A^{\ast\ast}\simeq 5$ (positive pulse) the final stable state is the AFM. Here the initial state is the same as in the previous figure (the slightly disturbed AFM). Again, the asymptotic, $t \gg 1$, picture does not depend on the initial state for $A\not\in(A^{\ast},A^{\ast\ast})$. The Landau-Lifshitz-Gilbert equation 
parameters are $\gamma = 1$, $\lambda = 1$, anisotropy parameter $\rho = 5$, the retarded exchange amplitude is $G = 10$.
	} \label{Fig4-5}
\end{figure*}

In the presence of the anisotropy, both states, the AFM and the FM, become stable. There exist two ways of switching the final destination of the system between these states.
The first way is changing the parameters of the dynamically frustrated potential. The example of such a switch by changing the characteristic frequency $\omega_0$ is shown in Fig.\,\ref{Fig4-5}a presenting the temporal evolution of quantity $\mathbf{{m_1}\cdot {m_1}}$, characterizing the state of the system.
The second possibility of the switching is the perturbation in a form of the single half-sinusoidal external magnetic field pulse. In this case, in contrast to the isotropic one, the switched state is stable. Depending on the direction of the pulse, the final stable state is either the FM or the AFM state. Figure\,\ref{Fig4-5}b
shows an example of such a switch.

The revealed behaviours are of a general character and maintain for a general case of the system with the arbitrary number of the magnetic moments. The behaviour of the exemplary four-site cluster is presented in the Supplementary Information (SI).

\section*{Discussion and conclusion}~~\\

We have studied spin system with retarded spin-spin interaction $J_{ij}$. This implies the non-Markovian type of the time-dependent magnetic interaction and leads to nontrivial dynamics of the interacting magnetic moments. The time-frustrated case where $J_{ij}(\omega=0)=\int_0^\infty J_{ij}(t)dt=0$ is the most interesting regime because in this case the sign of $J_{ij}(\omega=0)$ does not naively predict the arising at $t\to\infty$ magnetic configuration.

It is important to stress that the retardation causes the non-Hermiticity of the effective Hamiltonian of the interacting magnetic moments, therefore, the considered system is an effectively dissipative. Non-Hermitian quantum mechanics describing open dissipative systems is currently enjoying an intense explosive development\,\cite{Bender07_RoPiP,Znojil17_PRA,Leykam17_PRL,Bolduc16_Nc}, and further aspects and implications of the non-Hermitian behavior of the system in hand will be a subject of the forthcoming publication.

The retarded spin-spin interaction is realized in the systems with the superexchange where magnetic moments interact indirectly through a medium with the pronounced frequency dispersion, granular multiferroics offering an appealing example. In multiferroics, magnetic granules interact through a ferroelectric medium. Its polarization comprises several contributions with the different characteristic times, $\mathbf P=\mathbf P_{\rm el}+\mathbf P_{\rm ion}+\mathbf P_{\rm dipols} + \ldots$. Here the first ``elastic'' contribution is the polarization of the outer electron shells, the second one is related to the ion shifts, and the third contribution is related to dipole moments of molecules; the second and the third terms typically are responsible for the ferroelectricity. It is important that all the contributions except the first one are relatively slow, with their relaxation times being larger or of order of the inverse phonon frequencies for which 1\,THz is a natural scale\,\cite{Thornb67_JAP,Togo15_SM,Hinuma17_CMS}.
At the same time, $P_{\rm el}$ relaxation time is electronic, having the optical frequencies, being thus by several orders of magnitude shorter
($G$ in Eq.\,(\ref{Jt_2}) is obviously define by $P_{\rm el}$).
When magnetic moments evolve fast, the superexchange interaction involves only polarization of the outer electron shells, while slow evolution of magnetic moments involves the change of the ferroelectric polarization due to shift of ions. This is the picture in the frequency domain. In the time domain this physical mechanism leads to the retarded spin-spin interaction.

There are other frequency dependent exchange channels in the problem. We consider one of the most important cases. Other known mechanisms either look similar to the considered one or are suppressed by the Coulomb blockade \cite{Bruno95_PRB,Udalov18_AA}. Note here one must talk about the direct exchange modulated by the environment. We have avoided using this term, preferring a somewhat loose use of the word "superexchange". Moreover, in the exact sense of the word, there is no "complete" direct exchange in our case. At low frequencies, there is no overlap, it occurs only when the excitation of the medium is taken into account.

Let us also make a supporting remark. We have demonstrated the discussed effect for a particular set of parameters. A similar behavior of the system is observed while varying them in the wider range, in particular, when varying $\lambda$ by an order of magnitude up and down.

Mutliferroics, the systems with interacting magnetic and electric degrees of freedom, broaden the scope of the existing current
hardware concepts\,\cite{Huang20_NC,Noel20_N,Polshy20_N,Chen_SA,Manipa18_SA,Heron18_N,Leo18_N,Spaldi17_NRM,
Liu16_N,Mundy16_N,Mandal15_N,Doerr14_N,Farokh14_N,Heron14_N} and introduce the new ones, see\,\cite{omori_invitation_2019,CezeN_19,Lukyan21_NL,baudry_ferroelectric_2017}. Magnetic mutliferroic tunnel junctions promise the platform for the computers based on the non-binary (many-valued) logics\,\cite{Garcia2014Nature}. We demonstrated that owing to the dynamic frustration, the ferromagnetic and antiferromagnetic are stable and long-living states, hence having, in fact, the same energies.

In particular, the system that we have considered renders the tunnel junction {\it magnet-ferroelectric-magnet} with the time frustrated exchange having four different stable states, comprising two ferroelectric and two magnetic states.
Hence the computer element based on the TFM holds high potential for the four-valued logic hardware realizations.

As mentioned above, the absolute value of the damping frequency is in the terahertz range. It is defined by phonon characteristic and so is untunable. But it is possible to change in the wide range the time scale of the Landau-Lifshitz equation. Therefore, we can control
the FM-AFM switching via varying the $\gamma_0 M_s$, e.g. by diluting the magnetic granules, by pressure or other external influence. Thus, the optimal regime, when the most curious time-frustrated effects appear, can always be achieved.

\section*{Methods}
\subsection{Derivation of a general equation}
Taking magnetization as
\begin{equation}
\label{adj_def}
  \mathbf{m}_{i}(t)=\mathbf{m}^{0}_{i} + \mathbf{m}^{\delta}_{i}(t)\,,
\end{equation}
where $\mathbf{m}^{\delta}_{i}(t)$ is assumed small, one finds that
the corresponding linear approximation of the Landau-Lifshitz-Gilbert equation for $\mathbf{m}^{\delta}_{i}(t)$ assumes the form
\begin{gather}
\dot{\mathbf{m}}_{i}^{\delta }(t) = -\mathbf{b}_{i}^{\delta }(t)-\lambda
\lbrack {\mathbf{m}_{i}^{\delta }(t)\times }{\mathbf{b}_{i}^{0}]}-\lambda
\lbrack {\mathbf{m}_{i}^{0}\times }{\mathbf{b}_{i}^{\delta }(t)]} \label{LL_gen} \\
\mathbf{b}_{i}^{\delta }(t) = [{\mathbf{m}_{i}^{\delta }(t)\times }{\mathbf{h}_{i}^{0}}]+[{\mathbf{m}_{i}^{0}\times }{\mathbf{h}_{i}^{\delta }(t)]} \quad \quad \quad \quad \quad \label{LL_gen_2}\\
{\mathbf{h}_{i}^{\delta }(t)} = \sum_{NN}\int_{-\infty }^{t}J(t-\tau )
\mathbf{m}_{NN}^{\delta }(t)d\tau +\mathbf{h}^{\mathrm{ext}}(t) \label{LL_gen_3}\,.
\end{gather}
In the frustrated case with $\int_{-\infty}^{t \to \infty} J(t-\tau) d\tau = 0$, both $\mathbf{h}_{i}^{0}(t)=0$ and $\mathbf{b}_{i}^{0}(t)=0$ for large enough values of $t$, and Eq.\,(\ref{LL_gen}) reduces to Eq.\,(\ref{LL_frustr}) of the main text.
Taking the Fourier transform of (\ref{LL_frustr}), one gets
\begin{equation}
\label{LL_omega}
i\omega \mathbf{m}_{i}^{\delta}(\omega) = -\vecprod{\mathbf{m}_{i}^{0}}{\mathbf{h}_{i}^{\delta }(\omega)}
-\lambda\vecprod{\mathbf{m}_{i}^{0}}
{\vecprod{\mathbf{m}_{i}^{0}}{\mathbf{h}_{i}^{\delta }(\omega)}}\,,
\end{equation}
where
\begin{equation} \label{h_ss}
{\mathbf{h}_{i}^{\delta }}(\omega ) = \sum_{NN}J(\omega )\mathbf{m}_{NN}^{\delta }(\omega )+\mathbf{h}^{\mathrm{ext}}(\omega)\,.
\end{equation}
The double cross product in\,(\ref{LL_omega}) is
($\mathbf{m}_{i}^{0} \cdot \mathbf{m}_{i}^{0} = 1$)
\begin{equation}
\label{tri_vec}
\vecprod{\mathbf{m}_{i}^{0}}
{\vecprod{\mathbf{m}_{i}^{0}}{\mathbf{h}_{i}^{\delta }(\omega)}} =
\mathbf{m}_{i}^{0}\sclprod{\mathbf{m}_{i}^{0}}{\mathbf{h}_{i}^{\delta }(\omega)} - \mathbf{h}_{i}^{\delta }(\omega)\,,
\end{equation}
and the evolution equation for $i$-th for the Fourier transform of magnetic moment becomes Eq.\,(\ref{LL_omega1}) of the main text.

\subsection{Two-site cluster with the frustrated exchange}
For the two-site cluster the system of equations (\ref {LL_omega1}) for $\mathbf{m}_{1}^{\delta }(\omega )$, $\mathbf{m}_{2}^{\delta }(\omega )$ reads (we set $\mathbf{h}^{\mathrm{ext}}(\omega ) \parallel x$):
\begin{gather}
i\omega \mathbf{m}_{1}^{\delta }(\omega )=-[\mathbf{m}{_{1}^{0}\times
\mathbf{h}^{\mathrm{ext}}(\omega )]} - J(\omega )[{\mathbf{m}%
_{1}^{0}\times \mathbf{m}_{2}^{\delta }(\omega )]}  \notag \\
+\lambda \mathbf{h}^{\mathrm{ext}}(\omega ) + J(\omega )\lambda
\mathbf{m}_{2}^{\delta }(\omega ) \\
i\omega \mathbf{m}_{2}^{\delta }(\omega )=-[{\mathbf{m}_{2}^{0}\times
\mathbf{h}^{\mathrm{ext}}(\omega )]} - J(\omega )[{\mathbf{m}%
_{2}^{0}\times \mathbf{m}_{1}^{\delta }(\omega )]}  \notag \\
+\lambda \mathbf{h}^{\mathrm{ext}}(\omega ) + J(\omega )\lambda
\mathbf{m}_{1}^{\delta }(\omega )
\end{gather}
with (after projection to x- and y-axes) the determinant
\begin{equation}
\Delta(\omega) = \left(
\begin{array}{cccc} \label{det}
i\omega & 0 & -\lambda {J}(\omega) & - {J}(\omega)m_{1}^{0} \\
0 & i\omega & {J}(\omega)m_{1}^{0} & -\lambda {J}(\omega) \\
-\lambda {J}(\omega) & - {J}(\omega)m_{2}^{0} & i\omega & 0 \\
{J}(\omega)m_{2}^{0} & -\lambda {J}(\omega) & 0 & i\omega
\end{array}
\right)
\end{equation}
were
$m_{i}^{0} = \left\vert \mathbf{m}_{i}^{0}\right\vert $.

Taking the exchange in the form Eq.~\eqref{J_simp_2} that not only preserves the causality, but also $\mathrm{ Re\,} J(\omega)$ is even function of $\omega$ like $\mathrm{ Re\,}1/\epsilon(\omega)$. Then the characteristic equation
$\Delta(\omega) = 0$ has four roots
(apart from four stationary roots $\omega_{0} = 0$):
\begin{eqnarray}
\omega_{1,2} &=& -i\omega_0 \pm iG\sqrt{\left( \lambda +im_{1}^{0}\right) \left(\lambda +im_{2}^{0}\right) } \\
\omega_{3,4} &=& -i\omega_0 \pm iG\sqrt{\left( \lambda -im_{1}^{0}\right) \left(\lambda -im_{2}^{0}\right) }
\end{eqnarray}

For the FM case $m_{1}^{0}=m_{2}^{0}=+1$, we find
\begin{eqnarray}
\omega_{1,2} &=& -i\omega_0 \pm iG\left( \lambda +i\right)  \\
\omega_{3,4} &=& -i\omega_0 \pm iG\left( \lambda -i\right)
\end{eqnarray}
so the stability condition is $G\lambda < \omega_0 $.

For the AFM configuration $m_{1}^{0} = - m_{2}^{0} = +1$,
we have two double-degenerate roots
\begin{equation}
\omega_{1-4} = -i\omega_0 \pm iG \sqrt{1+\lambda^{2}}\,,
\end{equation}
and the resulting stability condition is
${G}\sqrt{1+\lambda^{2}} < \omega_0 $.

\bigskip
\noindent \textbf{References}
\vspace{-0.6cm}
\bibliographystyle{plain}
\bibliography{main}

\begin{thebibliography}{10}
\expandafter\ifx\csname url\endcsname\relax
  \def\url#1{\texttt{#1}}\fi
\expandafter\ifx\csname urlprefix\endcsname\relax\def\urlprefix{URL }\fi
\providecommand{\bibinfo}[2]{#2}
\providecommand{\eprint}[2][]{\url{#2}}

\bibitem{Magnetism2006}
\bibinfo{author}{St\"{o}r, J.} \& \bibinfo{author}{Siegmann, H.~C.}
\newblock \textit{\bibinfo{title}{Magnetism: {From} {Fundamentals} to
  {Nanoscale} {Dynamics}}}.
\newblock Springer {Series} in {Solid}-{State} {Sciences}
\newblock  (\bibinfo{publisher}{Springer-Verlag}, \bibinfo{address}{Berlin
  Heidelberg}, \bibinfo{year}{2006}).

\bibitem{Magnetism2007}
\bibinfo{author}{White, R.~M.}
\newblock \textit{\bibinfo{title}{Quantum {Theory} of {Magnetism}: {Magnetic}
  {Properties} of {Materials}}}.
\newblock Springer {Series} in {Solid}-{State} {Sciences}
  (\bibinfo{publisher}{Springer-Verlag}, \bibinfo{address}{Berlin Heidelberg},
  \bibinfo{year}{2007}),
\newblock \bibinfo{edition}{3} edn.

\bibitem{Magnetism2021}
\bibinfo{editor}{Coey, M.} \& \bibinfo{editor}{Parkin, S.} (eds.)
  \textit{\bibinfo{title}{Handbook of {Magnetism} and {Magnetic} {Materials}}}
\newblock  (\bibinfo{publisher}{Springer International Publishing},
  \bibinfo{year}{2021}).

\bibitem{Manipa19_N}
\bibinfo{author}{Manipatruni, S.} \textit{et~al.}
\newblock \bibinfo{title}{Scalable energy-efficient magnetoelectric
  spin–orbit logic}.
\newblock \textit{\bibinfo{journal}{Nature}} \textbf{\bibinfo{volume}{565}},
  \bibinfo{pages}{35--42}
\newblock  (\bibinfo{year}{2019}).

\bibitem{Kirilyuk10_RMP}
\bibinfo{author}{Kirilyuk, A.}, \bibinfo{author}{Kimel, A.~V.} \&
  \bibinfo{author}{Rasing, T.}
\newblock \bibinfo{title}{Ultrafast optical manipulation of magnetic order}.
\newblock \textit{\bibinfo{journal}{Rev. Mod. Phys.}}
  \textbf{\bibinfo{volume}{82}}, \bibinfo{pages}{2731--2784}
  (\bibinfo{year}{2010}).
\newblock \bibinfo{note}{Publisher: American Physical Society}.

\bibitem{Spaldi19_NM}
\bibinfo{author}{Spaldin, N.~A.} \& \bibinfo{author}{Ramesh, R.}
\newblock \bibinfo{title}{Advances in magnetoelectric multiferroics}.
\newblock \textit{\bibinfo{journal}{Nat. Mater.}}
  \textbf{\bibinfo{volume}{18}}, \bibinfo{pages}{203--212}
  (\bibinfo{year}{2019}).
\newblock \bibinfo{note}{Bandiera\_abtest: a Cg\_type: Nature Research Journals
  Number: 3 Primary\_atype: Reviews Publisher: Nature Publishing Group
  Subject\_term: Ferroelectrics and multiferroics;Magnetic properties and
  materials Subject\_term\_id:
  ferroelectrics-and-multiferroics;magnetic-properties-and-materials}.

\bibitem{Spaldi17_NRM}
\bibinfo{author}{Spaldin, N.~A.}
\newblock \bibinfo{title}{Multiferroics: from the cosmically large to the
  subatomically small}.
\newblock \textit{\bibinfo{journal}{Nat. Rev. Mater.}}
  \textbf{\bibinfo{volume}{2}}, \bibinfo{pages}{17017}
\newblock  (\bibinfo{year}{2017}).

\bibitem{Dong15_AP}
\bibinfo{author}{Dong, S.}, \bibinfo{author}{Liu, J.-M.},
  \bibinfo{author}{Cheong, S.-W.} \& \bibinfo{author}{Ren, Z.}
\newblock \bibinfo{title}{Multiferroic materials and magnetoelectric physics:
  symmetry, entanglement, excitation, and topology}.
\newblock \textit{\bibinfo{journal}{Adv. Phys.}} \textbf{\bibinfo{volume}{64}},
  \bibinfo{pages}{519--626}
\newblock  (\bibinfo{year}{2015}).

\bibitem{Fiebig16_NRM}
\bibinfo{author}{Fiebig, M.}, \bibinfo{author}{Lottermoser, T.},
  \bibinfo{author}{Meier, D.} \& \bibinfo{author}{Trassin, M.}
\newblock \bibinfo{title}{The evolution of multiferroics}.
\newblock \textit{\bibinfo{journal}{Nat. Rev. Mater.}}
  \textbf{\bibinfo{volume}{1}}, \bibinfo{pages}{16046}
\newblock  (\bibinfo{year}{2016}).

\bibitem{Udalov14_PRB}
\bibinfo{author}{Udalov, O.~G.}, \bibinfo{author}{Chtchelkatchev, N.~M.},
  \bibinfo{author}{Glatz, A.} \& \bibinfo{author}{Beloborodov, I.~S.}
\newblock \bibinfo{title}{Interplay of coulomb blockade and ferroelectricity in
  nanosized granular materials}.
\newblock \textit{\bibinfo{journal}{Phys. Rev. B}}
  \textbf{\bibinfo{volume}{89}}, \bibinfo{pages}{054203}
\newblock  (\bibinfo{year}{2014}).

\bibitem{Udalov14_PRBb}
\bibinfo{author}{Udalov, O.~G.}, \bibinfo{author}{Chtchelkatchev, N.~M.} \&
  \bibinfo{author}{Beloborodov, I.~S.}
\newblock \bibinfo{title}{Proximity coupling of a granular film with a
  ferroelectric substrate and giant electroresistance effect}.
\newblock \textit{\bibinfo{journal}{Phys. Rev. B}}
  \textbf{\bibinfo{volume}{90}}, \bibinfo{pages}{054201}
\newblock  (\bibinfo{year}{2014}).

\bibitem{Maity20_PSSB}
\bibinfo{author}{Maity, A.}, \bibinfo{author}{Schwesig, S.},
  \bibinfo{author}{Ziegler, F.}, \bibinfo{author}{Sobolev, O.} \&
  \bibinfo{author}{Eckold, G.}
\newblock \bibinfo{title}{Magnons in the {Multiferroic} {Phase} of {Cupric}
  {Oxide}}.
\newblock \textit{\bibinfo{journal}{Phys. Status Solidi B}}
  \textbf{\bibinfo{volume}{257}}, \bibinfo{pages}{1900704}
\newblock  (\bibinfo{year}{2020}).

\bibitem{Ma20_PSSR}
\bibinfo{author}{Ma, X.} \textit{et~al.}
\newblock \bibinfo{title}{Tunable {Valley} {Polarization} in a {Multiferroic}
  {CuCrP2Te6} {Monolayer}}.
\newblock \textit{\bibinfo{journal}{Phys. Status Solidi R}}
  \textbf{\bibinfo{volume}{14}}, \bibinfo{pages}{2000008}
\newblock  (\bibinfo{year}{2020}).

\bibitem{Fedoro14_PRBa}
\bibinfo{author}{Fedorov, S.~A.}, \bibinfo{author}{Korolkov, A.~E.},
  \bibinfo{author}{Chtchelkatchev, N.~M.}, \bibinfo{author}{Udalov, O.~G.} \&
  \bibinfo{author}{Beloborodov, I.~S.}
\newblock \bibinfo{title}{Memory effect in a ferroelectric single-electron
  transistor: Violation of conductance periodicity in the gate voltage}.
\newblock \textit{\bibinfo{journal}{Phys. Rev. B}}
  \textbf{\bibinfo{volume}{90}}, \bibinfo{pages}{195111}
\newblock  (\bibinfo{year}{2014}).

\bibitem{Secchi13_APN}
\bibinfo{author}{Secchi, A.}, \bibinfo{author}{Brener, S.},
  \bibinfo{author}{Lichtenstein, A.~I.} \& \bibinfo{author}{Katsnelson, M.~I.}
\newblock \bibinfo{title}{Non-equilibrium magnetic interactions in strongly
  correlated systems}.
\newblock \textit{\bibinfo{journal}{Ann. Phys. (N.Y.)}}
  \textbf{\bibinfo{volume}{333}}, \bibinfo{pages}{221--271}
\newblock  (\bibinfo{year}{2013}).

\bibitem{Claass17_NC}
\bibinfo{author}{Claassen, M.}, \bibinfo{author}{Jiang, H.-C.},
  \bibinfo{author}{Moritz, B.} \& \bibinfo{author}{Devereaux, T.~P.}
\newblock \bibinfo{title}{Dynamical time-reversal symmetry breaking and
  photo-induced chiral spin liquids in frustrated {Mott} insulators}.
\newblock \textit{\bibinfo{journal}{Nat. Commun.}}
  \textbf{\bibinfo{volume}{8}}, \bibinfo{pages}{1192}
\newblock  (\bibinfo{year}{2017}).

\bibitem{Mentin17_JPCM}
\bibinfo{author}{Mentink, J.~H.}
\newblock \bibinfo{title}{Manipulating magnetism by ultrafast control of the
  exchange interaction}.
\newblock \textit{\bibinfo{journal}{J. Phys.: Condens. Matter}}
  \textbf{\bibinfo{volume}{29}}, \bibinfo{pages}{453001}
  (\bibinfo{year}{2017}).
\newblock \bibinfo{note}{Publisher: IOP Publishing}.

\bibitem{Liu18_PRL}
\bibinfo{author}{Liu, J.}, \bibinfo{author}{Hejazi, K.} \&
  \bibinfo{author}{Balents, L.}
\newblock \bibinfo{title}{Floquet {Engineering} of {Multiorbital} {Mott}
  {Insulators}: {Applications} to {Orthorhombic} {Titanates}}.
\newblock \textit{\bibinfo{journal}{Phys. Rev. Lett.}}
  \textbf{\bibinfo{volume}{121}}, \bibinfo{pages}{107201}
  (\bibinfo{year}{2018}).
\newblock \bibinfo{note}{Publisher: American Physical Society}.

\bibitem{Chaudh19_PRB}
\bibinfo{author}{Chaudhary, S.}, \bibinfo{author}{Hsieh, D.} \&
  \bibinfo{author}{Refael, G.}
\newblock \bibinfo{title}{Orbital {Floquet} engineering of exchange
  interactions in magnetic materials}.
\newblock \textit{\bibinfo{journal}{Phys. Rev. B}}
  \textbf{\bibinfo{volume}{100}}, \bibinfo{pages}{220403}
  (\bibinfo{year}{2019}).
\newblock \bibinfo{note}{Publisher: American Physical Society}.

\bibitem{Ke20_PRR}
\bibinfo{author}{Ke, M.}, \bibinfo{author}{Asmar, M.~M.} \&
  \bibinfo{author}{Tse, W.-K.}
\newblock \bibinfo{title}{Nonequilibrium {RKKY} interaction in irradiated
  graphene}.
\newblock \textit{\bibinfo{journal}{Phys. Rev. Res.}}
  \textbf{\bibinfo{volume}{2}}, \bibinfo{pages}{033228} (\bibinfo{year}{2020}).
\newblock \bibinfo{note}{Publisher: American Physical Society}.

\bibitem{Mikhay20_PRL}
\bibinfo{author}{Mikhaylovskiy, R.} \textit{et~al.}
\newblock \bibinfo{title}{Resonant {Pumping} of
  \$d{\textbackslash}mathrm\{{\textbackslash}ensuremath\{-\}\}d\$ {Crystal}
  {Field} {Electronic} {Transitions} as a {Mechanism} of {Ultrafast} {Optical}
  {Control} of the {Exchange} {Interactions} in {Iron} {Oxides}}.
\newblock \textit{\bibinfo{journal}{Phys. Rev. Lett.}}
  \textbf{\bibinfo{volume}{125}}, \bibinfo{pages}{157201}
  (\bibinfo{year}{2020}).
\newblock \bibinfo{note}{Publisher: American Physical Society}.

\bibitem{Losada19_PRB}
\bibinfo{author}{Losada, J.~M.}, \bibinfo{author}{Brataas, A.} \&
  \bibinfo{author}{Qaiumzadeh, A.}
\newblock \bibinfo{title}{Ultrafast control of spin interactions in honeycomb
  antiferromagnetic insulators}.
\newblock \textit{\bibinfo{journal}{Phys. Rev. B}}
  \textbf{\bibinfo{volume}{100}}, \bibinfo{pages}{060410}
  (\bibinfo{year}{2019}).
\newblock \bibinfo{note}{Publisher: American Physical Society}.

\bibitem{Ron20_PRL}
\bibinfo{author}{Ron, A.} \textit{et~al.}
\newblock \bibinfo{title}{Ultrafast {Enhancement} of {Ferromagnetic} {Spin}
  {Exchange} {Induced} by {Ligand}-to-{Metal} {Charge} {Transfer}}.
\newblock \textit{\bibinfo{journal}{Phys. Rev. Lett.}}
  \textbf{\bibinfo{volume}{125}}, \bibinfo{pages}{197203}
  (\bibinfo{year}{2020}).
\newblock \bibinfo{note}{Publisher: American Physical Society}.

\bibitem{Udalov14_PRBa}
\bibinfo{author}{Udalov, O.~G.}, \bibinfo{author}{Chtchelkatchev, N.~M.} \&
  \bibinfo{author}{Beloborodov, I.~S.}
\newblock \bibinfo{title}{Coupling of ferroelectricity and ferromagnetism
  through coulomb blockade in composite multiferroics}.
\newblock \textit{\bibinfo{journal}{Phys. Rev. B}}
  \textbf{\bibinfo{volume}{89}}, \bibinfo{pages}{174203}
\newblock  (\bibinfo{year}{2014}).

\bibitem{Fedoro15_PRB}
\bibinfo{author}{Fedorov, S.~A.}, \bibinfo{author}{Chtchelkatchev, N.~M.},
  \bibinfo{author}{Udalov, O.~G.} \& \bibinfo{author}{Beloborodov, I.~S.}
\newblock \bibinfo{title}{Single-electron tunneling with slow insulators}.
\newblock \textit{\bibinfo{journal}{Phys. Rev. B}}
  \textbf{\bibinfo{volume}{92}}, \bibinfo{pages}{115425}
\newblock  (\bibinfo{year}{2015}).

\bibitem{landau2013electrodynamics}
\bibinfo{author}{Landau, L.~D.} \textit{et~al.}
\newblock \textit{\bibinfo{title}{Electrodynamics of continuous media}},
  vol.~\bibinfo{volume}{8}
\newblock  (\bibinfo{publisher}{Elsevier}, \bibinfo{year}{2013}).

\bibitem{Nikolic2019PRB}
\bibinfo{author}{Bajpai, U.} \& \bibinfo{author}{Nikoli\ifmmode~\acute{c}\else
  \'{c}\fi{}, B.~K.}
\newblock \bibinfo{title}{Time-retarded damping and magnetic inertia in the
  landau-lifshitz-gilbert equation self-consistently coupled to electronic
  time-dependent nonequilibrium green functions}.
\newblock \textit{\bibinfo{journal}{Phys. Rev. B}}
  \textbf{\bibinfo{volume}{99}}, \bibinfo{pages}{134409}
\newblock  (\bibinfo{year}{2019}).

\bibitem{John17_SR}
\bibinfo{author}{John, R.} \textit{et~al.}
\newblock \bibinfo{title}{Magnetisation switching of {FePt} nanoparticle
  recording medium by femtosecond laser pulses}.
\newblock \textit{\bibinfo{journal}{Sci. Rep.}} \textbf{\bibinfo{volume}{7}},
  \bibinfo{pages}{4114} (\bibinfo{year}{2017}).
\newblock \bibinfo{note}{Bandiera\_abtest: a Cc\_license\_type: cc\_by
  Cg\_type: Nature Research Journals Number: 1 Primary\_atype: Research
  Publisher: Nature Publishing Group Subject\_term: Magnetic properties and
  materials;Magneto-optics Subject\_term\_id:
  magnetic-properties-and-materials;magneto-optics}.

\bibitem{Liu17_APL}
\bibinfo{author}{Liu, Z.}, \bibinfo{author}{Huang, P.-W.}, \bibinfo{author}{Ju,
  G.} \& \bibinfo{author}{Victora, R.~H.}
\newblock \bibinfo{title}{Thermal switching probability distribution of {L10}
  {FePt} for heat assisted magnetic recording}.
\newblock \textit{\bibinfo{journal}{Appl. Phys. Lett.}}
  \textbf{\bibinfo{volume}{110}}, \bibinfo{pages}{182405}
  (\bibinfo{year}{2017}).
\newblock \bibinfo{note}{Publisher: American Institute of Physics}.

\bibitem{Jin18_PRE}
\bibinfo{author}{Jin, M.~H.}, \bibinfo{author}{Zheng, B.},
  \bibinfo{author}{Xiong, L.}, \bibinfo{author}{Zhou, N.~J.} \&
  \bibinfo{author}{Wang, L.}
\newblock \bibinfo{title}{Numerical simulations of critical dynamics in
  anisotropic magnetic films with the stochastic {Landau}-{Lifshitz}-{Gilbert}
  equation}.
\newblock \textit{\bibinfo{journal}{Phys. Rev. E}}
  \textbf{\bibinfo{volume}{98}}, \bibinfo{pages}{022126}
  (\bibinfo{year}{2018}).
\newblock \bibinfo{note}{Publisher: American Physical Society}.

\bibitem{Akbash15_JMMM}
\bibinfo{author}{Akbashev, A.}, \bibinfo{author}{Telegin, A.},
  \bibinfo{author}{Kaul, A.} \& \bibinfo{author}{Sukhorukov, Y.}
\newblock \bibinfo{title}{Granular and layered ferroelectric–ferromagnetic
  thin-film nanocomposites as promising materials with high magnetotransmission
  effect}.
\newblock \textit{\bibinfo{journal}{J. Magn. Magn. Mater.}}
  \textbf{\bibinfo{volume}{384}}
\newblock  (\bibinfo{year}{2015}).

\bibitem{Colla99_ASS}
\bibinfo{author}{Colla, E.} \textit{et~al.}
\newblock \bibinfo{title}{Ferroelectric phase transitions in materials embedded
  in porous media}.
\newblock \textit{\bibinfo{journal}{Acs. Sym. Ser.}}
  \textbf{\bibinfo{volume}{12}}, \bibinfo{pages}{963--966}
\newblock  (\bibinfo{year}{1999}).

\bibitem{Park04_PRL}
\bibinfo{author}{Park, S.}, \bibinfo{author}{Hur, N.}, \bibinfo{author}{Guha,
  S.} \& \bibinfo{author}{Cheong, S.-W.}
\newblock \bibinfo{title}{Percolative conduction in the
  half-metallic-ferromagnetic and ferroelectric mixture of
  ($\mathrm{La,Lu,Sr})$ $\mathrm{MnO}_3$}.
\newblock \textit{\bibinfo{journal}{Phys. Rev. Lett.}}
  \textbf{\bibinfo{volume}{92}}
\newblock  (\bibinfo{year}{2004}).

\bibitem{Mandal06_PRB}
\bibinfo{author}{Mandal, P.}, \bibinfo{author}{Choudhury, P.} \&
  \bibinfo{author}{Ghosh, B.}
\newblock \bibinfo{title}{Electronic transport in ferroelectric-ferromagnetic
  composites $\mathrm{La}_{5/8}$ $(\mathrm{Ba}, \mathrm{Ca})_{3/8}$
  $\mathrm{MnO}_{3}$: $\mathrm{LuMnO}_{3}$}.
\newblock \textit{\bibinfo{journal}{Phys. Rev. B}}
  \textbf{\bibinfo{volume}{74}}
\newblock  (\bibinfo{year}{2006}).

\bibitem{Bender07_RoPiP}
\bibinfo{author}{Bender, C.~M.}
\newblock \bibinfo{title}{Making sense of non-hermitian hamiltonians}.
\newblock \textit{\bibinfo{journal}{Reports on Progress in Physics}}
  \textbf{\bibinfo{volume}{70}}, \bibinfo{pages}{947}
\newblock  (\bibinfo{year}{2007}).

\bibitem{Znojil17_PRA}
\bibinfo{author}{Znojil, M.}, \bibinfo{author}{Semor\'adov\'a, I.},
  \bibinfo{author}{R\ifmmode \mathring{u}\else \r{u}\fi{}\ifmmode
  \check{z}\else \v{z}\fi{}i\ifmmode~\check{c}\else \v{c}\fi{}ka, F. c.~v.},
  \bibinfo{author}{Moulla, H.} \& \bibinfo{author}{Leghrib, I.}
\newblock \bibinfo{title}{Problem of the coexistence of several non-hermitian
  observables in $\mathcal{PT}$-symmetric quantum mechanics}.
\newblock \textit{\bibinfo{journal}{Phys. Rev. A}}
  \textbf{\bibinfo{volume}{95}}, \bibinfo{pages}{042122}
\newblock  (\bibinfo{year}{2017}).

\bibitem{Leykam17_PRL}
\bibinfo{author}{Leykam, D.}, \bibinfo{author}{Bliokh, K.~Y.},
  \bibinfo{author}{Huang, C.}, \bibinfo{author}{Chong, Y.} \&
  \bibinfo{author}{Nori, F.}
\newblock \bibinfo{title}{Edge {Modes}, {Degeneracies}, and {Topological}
  {Numbers} in {Non}-{Hermitian} {Systems}}.
\newblock \textit{\bibinfo{journal}{Physical Review Letters}}
  \textbf{\bibinfo{volume}{118}}, \bibinfo{pages}{040401}
  (\bibinfo{year}{2017}).
\newblock \bibinfo{note}{Publisher: American Physical Society}.

\bibitem{Bolduc16_Nc}
\bibinfo{author}{Bolduc, E.}, \bibinfo{author}{Gariepy, G.} \&
  \bibinfo{author}{Leach, J.}
\newblock \bibinfo{title}{Direct measurement of large-scale quantum states via
  expectation values of non-hermitian matrices}.
\newblock \textit{\bibinfo{journal}{Nat. Commun.}}
  \textbf{\bibinfo{volume}{7}}, \bibinfo{pages}{10439}
\newblock  (\bibinfo{year}{2016}).

\bibitem{Thornb67_JAP}
\bibinfo{author}{Thornber, K.~K.}, \bibinfo{author}{McGill, T.~C.} \&
  \bibinfo{author}{Mead, C.~A.}
\newblock \bibinfo{title}{The {Tunneling} {Time} of an {Electron}}.
\newblock \textit{\bibinfo{journal}{J. Appl. Phys.}}
  \textbf{\bibinfo{volume}{38}}, \bibinfo{pages}{2384--2385}
  (\bibinfo{year}{1967}).
\newblock \bibinfo{note}{Publisher: American Institute of Physics}.

\bibitem{Togo15_SM}
\bibinfo{author}{Togo, A.} \& \bibinfo{author}{Tanaka, I.}
\newblock \bibinfo{title}{First principles phonon calculations in materials
  science}.
\newblock \textit{\bibinfo{journal}{Scr. Mater.}}
  \textbf{\bibinfo{volume}{108}}, \bibinfo{pages}{1--5}
\newblock  (\bibinfo{year}{2015}).

\bibitem{Hinuma17_CMS}
\bibinfo{author}{Hinuma, Y.}, \bibinfo{author}{Pizzi, G.},
  \bibinfo{author}{Kumagai, Y.}, \bibinfo{author}{Oba, F.} \&
  \bibinfo{author}{Tanaka, I.}
\newblock \bibinfo{title}{Band structure diagram paths based on
  crystallography}.
\newblock \textit{\bibinfo{journal}{Comput. Mater. Sci.}}
  \textbf{\bibinfo{volume}{128}}, \bibinfo{pages}{140--184}
\newblock  (\bibinfo{year}{2017}).

\bibitem{Bruno95_PRB}
\bibinfo{author}{Bruno, P.}
\newblock \bibinfo{title}{Theory of interlayer magnetic coupling}.
\newblock \textit{\bibinfo{journal}{Phys. Rev. B}}
  \textbf{\bibinfo{volume}{52}}, \bibinfo{pages}{411--439}
  (\bibinfo{year}{1995}).
\newblock \bibinfo{note}{Publisher: American Physical Society}.

\bibitem{Udalov18_AA}
\bibinfo{author}{Udalov, O.~G.} \& \bibinfo{author}{Beloborodov, I.~S.}
\newblock \bibinfo{title}{The {Coulomb} based magneto-electric coupling in
  multiferroic tunnel junctions and granular multiferroics}.
\newblock \textit{\bibinfo{journal}{AIP Adv.}} \textbf{\bibinfo{volume}{8}},
  \bibinfo{pages}{055810} (\bibinfo{year}{2018}).
\newblock \bibinfo{note}{Publisher: American Institute of Physics}.

\bibitem{Huang20_NC}
\bibinfo{author}{Huang, Y.-L.} \textit{et~al.}
\newblock \bibinfo{title}{Manipulating magnetoelectric energy landscape in
  multiferroics}.
\newblock \textit{\bibinfo{journal}{Nat. Commun.}}
  \textbf{\bibinfo{volume}{11}}, \bibinfo{pages}{2836} (\bibinfo{year}{2020}).
\newblock \bibinfo{note}{Bandiera\_abtest: a Cc\_license\_type: cc\_by
  Cg\_type: Nature Research Journals Number: 1 Primary\_atype: Research
  Publisher: Nature Publishing Group Subject\_term: Condensed-matter
  physics;Nanoscale materials Subject\_term\_id:
  condensed-matter-physics;nanoscale-materials}.

\bibitem{Noel20_N}
\bibinfo{author}{Noël, P.} \textit{et~al.}
\newblock \bibinfo{title}{Non-volatile electric control of spin–charge
  conversion in a srtio3 rashba system}.
\newblock \textit{\bibinfo{journal}{Nature}} \textbf{\bibinfo{volume}{580}},
  \bibinfo{pages}{483--486}
\newblock  (\bibinfo{year}{2020}).

\bibitem{Polshy20_N}
\bibinfo{author}{Polshyn, H.} \textit{et~al.}
\newblock \bibinfo{title}{Electrical switching of magnetic order in an orbital
  chern insulator}.
\newblock \textit{\bibinfo{journal}{Nature}} \textbf{\bibinfo{volume}{588}},
  \bibinfo{pages}{1--5}
\newblock  (\bibinfo{year}{2020}).

\bibitem{Chen_SA}
\bibinfo{author}{Chen, A.} \textit{et~al.}
\newblock \bibinfo{title}{Full voltage manipulation of the resistance of a
  magnetic tunnel junction}.
\newblock \textit{\bibinfo{journal}{Sci. Adv.}} \textbf{\bibinfo{volume}{5}},
  \bibinfo{pages}{eaay5141} (\bibinfo{year}{2019}).
\newblock \bibinfo{note}{Publisher: American Association for the Advancement of
  Science}.

\bibitem{Manipa18_SA}
\bibinfo{author}{Manipatruni, S.} \textit{et~al.}
\newblock \bibinfo{title}{Voltage control of unidirectional anisotropy in
  ferromagnet-multiferroic system}.
\newblock \textit{\bibinfo{journal}{Sci. Adv.}} \textbf{\bibinfo{volume}{4}},
  \bibinfo{pages}{eaat4229} (\bibinfo{year}{2018}).
\newblock \bibinfo{note}{Publisher: American Association for the Advancement of
  Science}.

\bibitem{Heron18_N}
\bibinfo{author}{Heron, J.} \& \bibinfo{author}{Mundy, J.}
\newblock \bibinfo{title}{Electric and magnetic domains inverted by a magnetic
  field}.
\newblock \textit{\bibinfo{journal}{Nature}} \textbf{\bibinfo{volume}{560}},
  \bibinfo{pages}{435--436}
\newblock  (\bibinfo{year}{2018}).

\bibitem{Leo18_N}
\bibinfo{author}{Leo, N.} \textit{et~al.}
\newblock \bibinfo{title}{Magnetoelectric inversion of domain patterns}.
\newblock \textit{\bibinfo{journal}{Nature}} \textbf{\bibinfo{volume}{560}},
  \bibinfo{pages}{466--470}
\newblock  (\bibinfo{year}{2018}).

\bibitem{Liu16_N}
\bibinfo{author}{Liu, S.}, \bibinfo{author}{Grinberg, I.} \&
  \bibinfo{author}{Rappe, A.}
\newblock \bibinfo{title}{Intrinsic ferroelectric switching from first
  principles}.
\newblock \textit{\bibinfo{journal}{Nature}} \textbf{\bibinfo{volume}{534}},
  \bibinfo{pages}{360--363}
\newblock  (\bibinfo{year}{2016}).

\bibitem{Mundy16_N}
\bibinfo{author}{Mundy, J.} \textit{et~al.}
\newblock \bibinfo{title}{Atomically engineered ferroic layers yield a
  room-temperature magnetoelectric multiferroic}.
\newblock \textit{\bibinfo{journal}{Nature}} \textbf{\bibinfo{volume}{537}},
  \bibinfo{pages}{523--527}
\newblock  (\bibinfo{year}{2016}).

\bibitem{Mandal15_N}
\bibinfo{author}{Mandal, P.} \textit{et~al.}
\newblock \bibinfo{title}{Designing switchable polarization and magnetization
  at room temperature in an oxide}.
\newblock \textit{\bibinfo{journal}{Nature}} \textbf{\bibinfo{volume}{525}},
  \bibinfo{pages}{363 -- 366}
\newblock  (\bibinfo{year}{2015}).

\bibitem{Doerr14_N}
\bibinfo{author}{Dörr, K.} \& \bibinfo{author}{Herklotz, A.}
\newblock \bibinfo{title}{Materials science: Two steps for a magnetoelectric
  switch}.
\newblock \textit{\bibinfo{journal}{Nature}} \textbf{\bibinfo{volume}{516}},
  \bibinfo{pages}{337--8}
\newblock  (\bibinfo{year}{2014}).

\bibitem{Farokh14_N}
\bibinfo{author}{Farokhipoor, S.} \textit{et~al.}
\newblock \bibinfo{title}{Artificial chemical and magnetic structure at the
  domain walls of an epitaxial oxide}.
\newblock \textit{\bibinfo{journal}{Nature}} \textbf{\bibinfo{volume}{515}},
  \bibinfo{pages}{379}
\newblock  (\bibinfo{year}{2014}).

\bibitem{Heron14_N}
\bibinfo{author}{Heron, J.} \textit{et~al.}
\newblock \bibinfo{title}{Deterministic switching of ferromagnetism at room
  temperature using an electric field}.
\newblock \textit{\bibinfo{journal}{Nature}} \textbf{\bibinfo{volume}{516}},
  \bibinfo{pages}{370--373}
\newblock  (\bibinfo{year}{2014}).

\bibitem{omori_invitation_2019}
\bibinfo{author}{Omori, H.} \& \bibinfo{author}{Wansing, H.}
\newblock \bibinfo{title}{An {Invitation} to {New} {Essays} on {Belnap}-{Dunn}
  {Logic}}.
\newblock In \bibinfo{editor}{Omori, H.} \& \bibinfo{editor}{Wansing, H.}
  (eds.) \textit{\bibinfo{booktitle}{New {Essays} on {Belnap}-­{Dunn}
  {Logic}}}, Synthese {Library}, \bibinfo{pages}{1--9}
\newblock  (\bibinfo{publisher}{Springer International Publishing},
  \bibinfo{address}{Cham}, \bibinfo{year}{2019}).

\bibitem{CezeN_19}
\bibinfo{author}{Ceze, L.}, \bibinfo{author}{Nivala, J.} \&
  \bibinfo{author}{Strauss, K.}
\newblock \bibinfo{title}{Molecular digital data storage using dna}.
\newblock \textit{\bibinfo{journal}{Nat. Rev. Genet.}}
  \textbf{\bibinfo{volume}{20}}
\newblock  (\bibinfo{year}{2019}).

\bibitem{Lukyan21_NL}
\bibinfo{author}{Lukyanchuk, I.} \textit{et~al.}
\newblock \bibinfo{title}{High-symmetry polarization domains in low-symmetry
  ferroelectrics}.
\newblock \textit{\bibinfo{journal}{Nano letters}}
  \textbf{\bibinfo{volume}{14}}
\newblock  (\bibinfo{year}{2013}).

\bibitem{baudry_ferroelectric_2017}
\bibinfo{author}{Baudry, L.}, \bibinfo{author}{Lukyanchuk, I.} \&
  \bibinfo{author}{Vinokur, V.~M.}
\newblock \bibinfo{title}{Ferroelectric symmetry-protected multibit memory
  cell}.
\newblock \textit{\bibinfo{journal}{Sci. Rep.}} \textbf{\bibinfo{volume}{7}},
  \bibinfo{pages}{42196}
\newblock  (\bibinfo{year}{2017}).

\bibitem{Garcia2014Nature}
\bibinfo{author}{Garcia, V.} \& \bibinfo{author}{Bibes, M.}
\newblock \bibinfo{title}{Ferroelectric tunnel junctions for information
  storage and processing}.
\newblock \textit{\bibinfo{journal}{Nat. Commun.}}
  \textbf{\bibinfo{volume}{5}}, \bibinfo{pages}{4289}
\newblock  (\bibinfo{year}{2014}).

\end{thebibliography}

\vskip 12pt	

\section*{Acknowledgements}~~\newline
The work of V.M.V was supported by Terra Quantum AG.
The work of N.M.Ch. was supported by the Russian Science Foundation (grant 18-12-00438).
The work of V.E.V. and A.V.M. was supported by the Russian Foundation for Basic Research  (grant 19-02-00509 A).

\section*{Author contribution}~~\newline
N.M.C and A.V.M. conceived the work and performed calculations, V.E.V. carried out numerical simulations, V.M.V. took part in calculations and outlining the project, N.M.C, A.V.M., and V.M.V. lead the interpretation of the results and wrote the manuscript, all authors discussed the manuscript.

\section*{Competing interests}~~\newline
 The authors declare  no competing interests.

\section*{Additional information}~~\newline
\noindent \textbf{Supplementary information} is \,available\,in\,the\,online\,version\,of\,the\,paper.

\bigskip

\noindent \textbf{Correspondence}\,should\,be\,addressed\,to\,V.M.V.\,(vmvinokour@gmail.com).

\end{document}